\documentclass[conference]{IEEEtran}
\IEEEoverridecommandlockouts
\usepackage{balance}
\usepackage{booktabs}
\usepackage[table,xcdraw]{xcolor}
\usepackage{listings}

\usepackage{cite}
\usepackage{amsmath,amssymb,amsfonts}
\usepackage{algorithmic}
\usepackage{graphicx}
\usepackage{textcomp}
\usepackage{xcolor}
\usepackage{authblk}
\usepackage{multirow}
\def\BibTeX{{\rm B\kern-.05em{\sc i\kern-.025em b}\kern-.08em
    T\kern-.1667em\lower.7ex\hbox{E}\kern-.125emX}}
\begin{document}
\title{IntrinTrans: LLM-based Intrinsic Code Translator for RISC-V Vector}

\author[1,2]{Liutong Han}
\author[1,2]{Zhiyuan Tan}
\author[1]{Hongbin Zhang}
\author[3]{Pengcheng Wang\textsuperscript{\textdagger}\thanks{\textsuperscript{\textdagger}Corresponding author: Mingjie Xing (mingjie@iscas.ac.cn) and Pengcheng Wang (wangpengcheng.pp@bytedance.com)}}
\author[1,2]{Chu Kang}
\author[1]{Mingjie Xing\textsuperscript{\textdagger}}
\author[1]{Yanjun Wu}

\affil[1]{Institute of Software, Chinese Academy of Sciences, Beijing, China}
\affil[2]{University of Chinese Academy of Sciences, Beijing, China}
\affil[3]{ByteDance, Beijing, China}

\maketitle

\begin{abstract}
The use of intrinsic functions to leverage hardware-specific capabilities is a crucial approach for optimizing library performance. Many mainstream libraries implement a large number of vectorized algorithms on Arm or x86 SIMD (Single-Instruction, Multiple-Data) intrinsic functions. Translating existing vectorized intrinsic code into the intrinsics of an emerging architecture is a practical and effective approach. However, current cross-architecture translation largely relies on manual rewriting or rule-based mapping methods, which are both time-consuming and prone to errors. We present \texttt{IntrinTrans}, a LLM-based agent that utilizes compile-and-test feedback to translate intrinsic code across architectures automatically, and further optimizes the generated intrinsics using register-usage information derived from liveness analysis. To evaluate the effectiveness of our method, we used \texttt{IntrinTrans} to translate the open-source benchmark from Arm Neon Intrinsic to the emerging RISC-V Vector (RVV) Intrinsic implementation and compared its performance with that of the native RVV implementation. Our experiments show that advanced LLMs can generate semantically correct RVV Intrinsic functions with only a finite number of iterations. Depending on the base LLMs, the pass rate ranges from 47\% to 100\%, achieving performance similar to the native implementation ($0.85\times$ to $1.28\times$).

\end{abstract}


\section{Introduction}

Data-Level Parallelism (DLP) is a cornerstone of modern high-performance computing, exploited via Single-Instruction, Multiple-Data (SIMD) extensions in processors. Architectures like Intel x86 SSE or AVX\cite{intelIntelAdvanced} and Arm Neon\cite{ArmNeon} or SVE\cite{armsve} provide robust SIMD capabilities. The RISC-V open standard is rapidly gaining traction, with its Vector (RVV) extension offering a flexible and powerful approach to DLP.

To leverage these capabilities, developers often use intrinsic functions, a set of compiler-supported functions that map directly to specific SIMD instructions. This approach strikes a critical balance, offering near-assembly-level performance control from within a high-level language like C/C++. Consequently, intrinsics are pervasive in performance-critical libraries.

However, this performance improvement comes at a significant development cost because the intrinsic functions are inherently non-portable. Each ISA has its own intrinsic functions, leading to substantial redundant effort when porting a library to a new architecture. This challenge is particularly acute for the burgeoning RISC-V ecosystem. Migrating the vast existing body of vectorized code from Arm or x86 to RVV is a significant bottleneck. Existing rule-based translation tools (e.g., SSE2RVV\cite{sse2rvv}, Neon2RVV\cite{neon2rvv}) attempt to address this by mapping source intrinsics to target equivalents, but the unique features of RVV, like the sizeless types and dynamic vector-length execution model, have no direct equivalents in fixed-length SIMD architectures, causing these direct mappings to fail in preserving both correctness and performance.

To address these challenges, we introduce \texttt{IntrinTrans}, a LLM-based agent framework for translating vectorized Arm Neon intrinsics code to RISC-V Vector intrinsics. We frame the translation task as a collaborative process handled by specialized agents with distinct roles: a Translator, a Compilation Executor, a Test Executor, and an Optimizer. The agents' interactions are orchestrated by a finite state machine (FSM), creating an iterative feedback loop where code is translated, compiled, tested for correctness, and then optimized for performance using feedback from static liveness analysis.

To evaluate our method, we used Arm Neon intrinsic code from VecIntrinBench \cite{vecIntrinBench} as source files input into \texttt{IntrinTrans} and evaluated the correctness and performance of its translation output. Experiments using nine different LLMs demonstrate that \texttt{IntrinTrans} can successfully translate complex Arm Neon inline functions into correct and high-performance RVV code, in some cases even outperforming expert-written implementations.

\section{Background and motivation}

\subsection{RISC-V Vector Extension}
% TODO 主要差异

% RISC-V 是一种新兴的开源精简指令集架构,其模块化的特性使得RISC-V架构能够更好地适应不同场景和需求下的处理器设计.RISC-V向量扩展是指令集的可选模块之一,旨在提供数据级并行能力,从而大幅提升数据处理效率.此扩展定义了一组向量指令及相关寄存器,从而允许程序使用向量运算部件对大量数据进行并行处理.李若时[17]等人评估了RISC-V向量扩展在提升计算机视觉算法效率方面的效果,指出性能提升可达2.98倍,展现了RISC-V向量扩展在高性能计算领域的应用前景.相较于其他支持数据级并行的指令集架构,RISC-V向量扩展具有如下特性:

RISC-V is an emerging open-source RISC architecture. Its modular design allows the RISC-V architecture to better adapt to diverse processor design requirements across various application domains. Among its optional extensions, the RISC-V Vector Extension (RVV) provides data-level parallelism, substantially improving computational efficiency for data-intensive workloads. This extension defines a set of vector instructions and corresponding registers, enabling programs to exploit vector processing units to operate on large datasets in parallel. Li et al.~\cite{JCST-2101-11266} evaluated the performance benefits of RVV on computer vision algorithms and reported up to a 2.98× speedup, demonstrating the strong potential of RVV in high-performance computing domains. The RVV extension features a design philosophy that significantly diverges from traditional fixed-length SIMD architectures like Arm Neon. Its two core hardware features are Vector Length Agnosticism (VLA) and vector register grouping (\texttt{LMUL}). VLA allows the same vectorized binary code to run efficiently on hardware with different vector register lengths (\texttt{VLEN}, e.g., 128-bit or 256-bit) without modification, thereby providing binary compatibility. Meanwhile, \texttt{LMUL} permits grouping multiple physical vector registers into a single, logically larger register set, enabling a single instruction to process more data elements and offering additional flexibility for performance tuning.

\begin{figure}[h]
  \centering
  \includegraphics[width=0.6\linewidth]{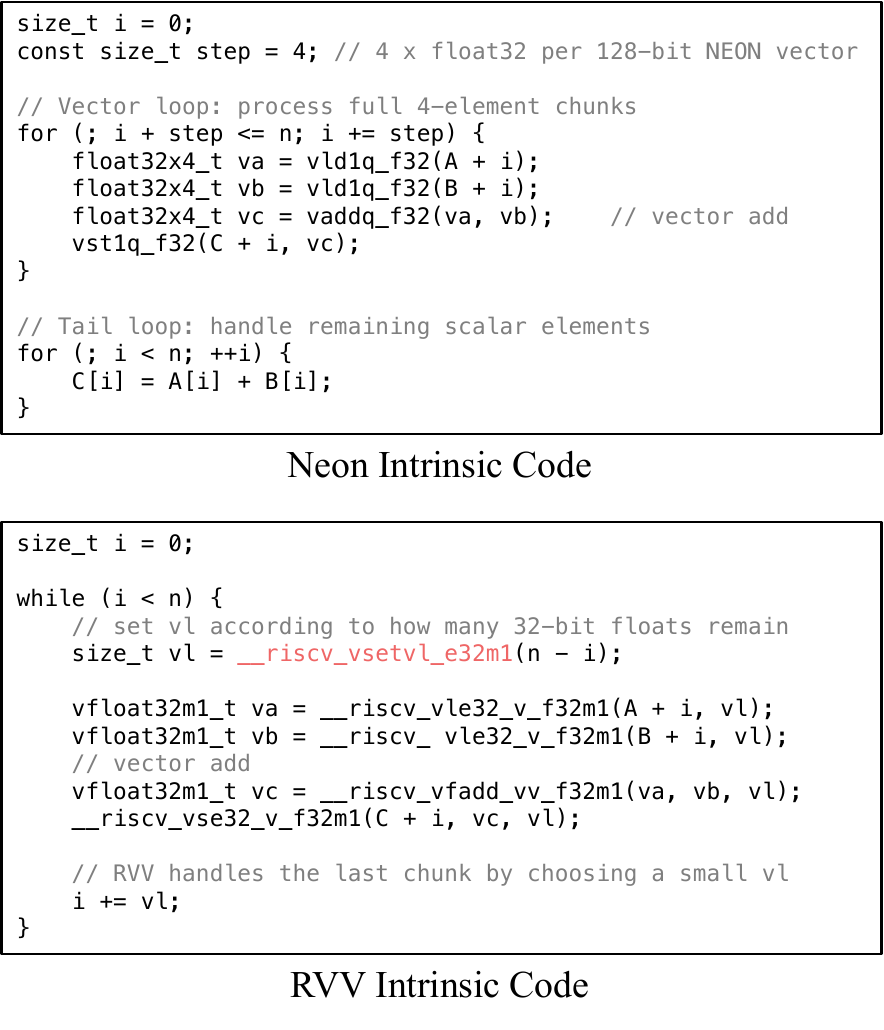}
  \caption{Comparison of Arm Neon and RISC-V Vector Addition}
  \label{fig:code-diff}
\end{figure}

These hardware features give rise to a programming model fundamentally different from that of fixed-length SIMD, centered on the introduction of the vector length register (\texttt{VL}). As illustrated in Figure \ref{fig:code-diff}, Arm Neon code must handle a fixed-size (e.g., 4-element) vector loop and requires an additional scalar tail loop to process any remaining data that does not fill a full vector. In contrast, RVV dynamically sets \texttt{VL} inside the loop by calling the \texttt{vsetvl} instruction based on the number of remaining elements. All subsequent vector instructions operate on \texttt{VL} elements, thus handling the loop tail implicitly within the main vector loop. This model eliminates the need for cumbersome and inefficient scalar tail-handling code, resulting in a more concise structure and higher execution efficiency, while also increasing the complexity of translating intrinsic code.

\subsection{Rule-based Intrinsic Translation}

% TODO：sse2rvv / Neon2RVV 头文件方案，成功率低，性能不佳。以Neon2RVV在几个代表性case上的预实验暴露出的成功率问题，且指出这种方案不能利用RISC-V Vector章节中提到的 VLA 和 LMUL 特性作为motivation

% 现有的面向RISC-V架构的代码翻译工作\cite{}，主要集中在基于规则映射（rule-based）的翻译策略上。该方法通过分析源架构与目标架构指令集之间的功能对应关系，构建指令映射表，从而在翻译过程中将源指令替换为功能等效或近似的目标指令。这种方法具有实现简单、便于扩展的优点，已在多个工具链与框架中得到应用，如XXX\cite{}。然而，由于不同指令集在语义、粒度和执行模型上的差异，简单的指令级映射往往难以保证翻译后的程序在性能上的一致性。此外，由于 RVV 扩展具有 VLA 以及 LMUL 等新特性，传统定长向量代码在翻译过程中无法利用RVV变长向量特性充分发挥性能。如图\ref{fig:performance-d}所示，在多种case上经由指令映射翻译的代码性能显著降低（提具体数据），难以满足高性能代码库的性能需求。因此这种方法具有较大的局限性。

Existing translation efforts targeting the RISC-V architecture\cite{neon2rvv,sse2rvv} primarily adopt rule-based translation strategies. Such approaches establish a mapping between the source and target instruction sets by analyzing their functional correspondences, thereby replacing source instructions with semantically equivalent or approximately equivalent target instructions during translation. This method is straightforward to implement and easy to extend, and has been widely integrated into several GitHub repos such as Dragonfly~\cite{dragonfly} and OpenXRay~\cite{openxray_xray16}.

However, due to inherent differences in instruction set semantics, granularity, and execution models, purely instruction-level mappings often fail to preserve performance consistency between the original and translated programs. Moreover, RVV introduces novel features such as VLA and LMUL, which are not directly compatible with traditional fixed-length vector architectures. Consequently, code translated from fixed-length vector ISAs cannot fully exploit RVV’s dynamic vector capabilities, leading to suboptimal utilization of hardware parallelism.

\begin{figure}[t]
  \centering
  \includegraphics[width=\linewidth]{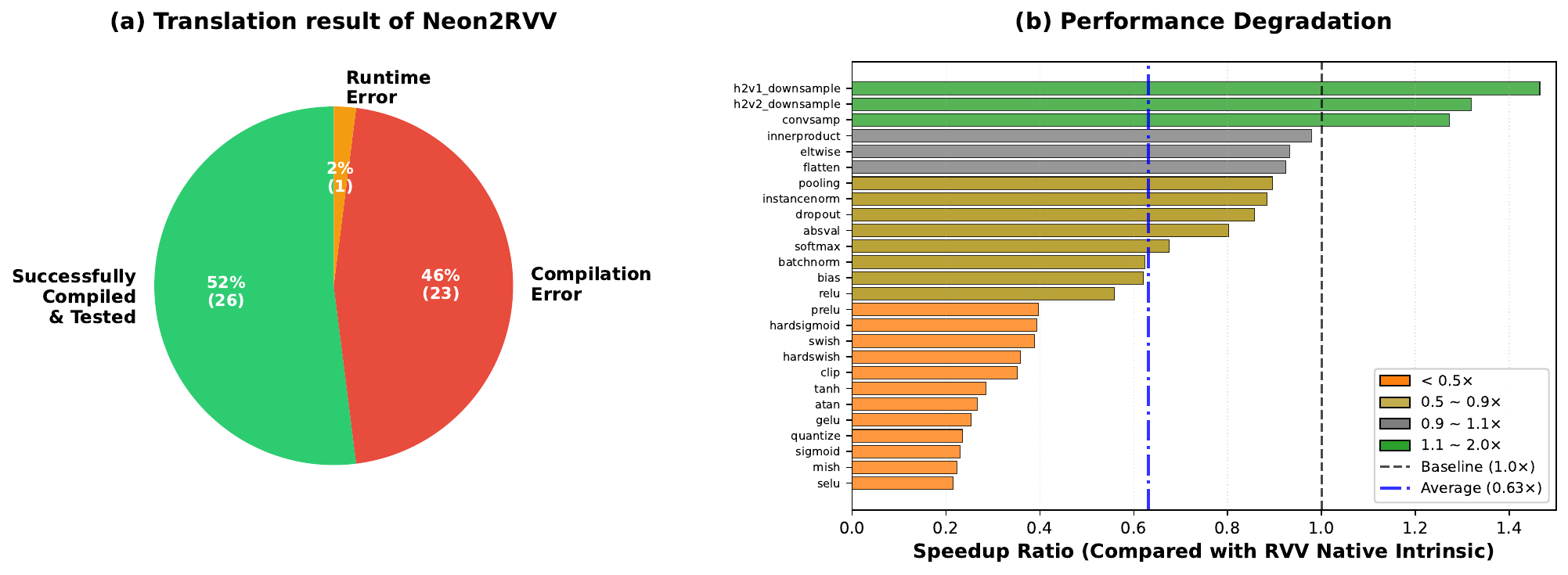}
  \caption{Correctness and Performance Evaluation of Neon2RVV Translation}
  \label{fig:performance-diff}
\end{figure}

As shown in Figure \ref{fig:performance-diff}, utilizing Neon2RVV achieves a compile success rate of only 52\%, indicating that the majority of translated vectorized functions fail to compile or pass testing. In terms of performance, the translated code exhibits a noticeable degradation across most benchmark cases, with an overall average speedup of merely $0.62\times$ compared to hand-optimized RVV implementations. Although a few cases achieve higher performance (discussed in Section \ref{sec:exp}), these results highlight that rule-based intrinsic translation approaches, while capable of ensuring partial functional correctness, struggle to deliver performance portability across architectures.

\subsection{LLM-based Code Translation}

Large Language Models (LLMs) have demonstrated remarkable capabilities in code generation~\cite{openai-o3,deepseek-r1}, showing significant potential to address the challenges above. The field of LLMs has advanced at a rapid pace, with notable recent improvements in coding proficiency exhibited by state-of-the-art models. Initial explorations~\cite{llm-vectorizer,vectrans} have begun to apply LLM-based agent systems to the vectorization process of scalar C code to x86 intrinsics. 

However, no existing work has systematically investigated the application of LLMs to the domain of cross-architecture intrinsic translation. Unlike traditional methods that rely on manually constructed instruction-mapping rules, LLMs can learn cross-language and cross-architecture semantic correspondences from large-scale corpora, thereby showing promise in handling complex translation tasks that lack one-to-one mappings. Some prior work \cite{CRT} has strengthened LLMs’ assembly-level translation ability via fine-tuning; however, both training and fine-tuning approaches incur significant GPU resource costs. Other studies \cite{UniTrans, UniTranslator} demonstrate that, through prompt engineering and interactive agent workflows, LLMs can achieve high-quality source-to-target code migration by leveraging contextual understanding and pattern induction even when explicit mapping rules are unavailable. Consequently, LLM-agent–based code translation represents a highly promising research avenue. 

Our previous work \cite{vecIntrinBench} proposed a benchmark for evaluating the migration capabilities of intrinsic code, which includes 50 function-level tasks from open-source repositories, implemented by RVV intrinsics and Arm Neon intrinsics, along with comprehensive functional and performance test cases. We also found that naive prompting strategies, which directly instruct an LLM to translate intrinsics from other architectures to RVV intrinsics, yield suboptimal results with low success rates.

\section{Design of IntrinTrans}

\begin{figure}[t]
  \centering
  \includegraphics[width=0.95\linewidth]{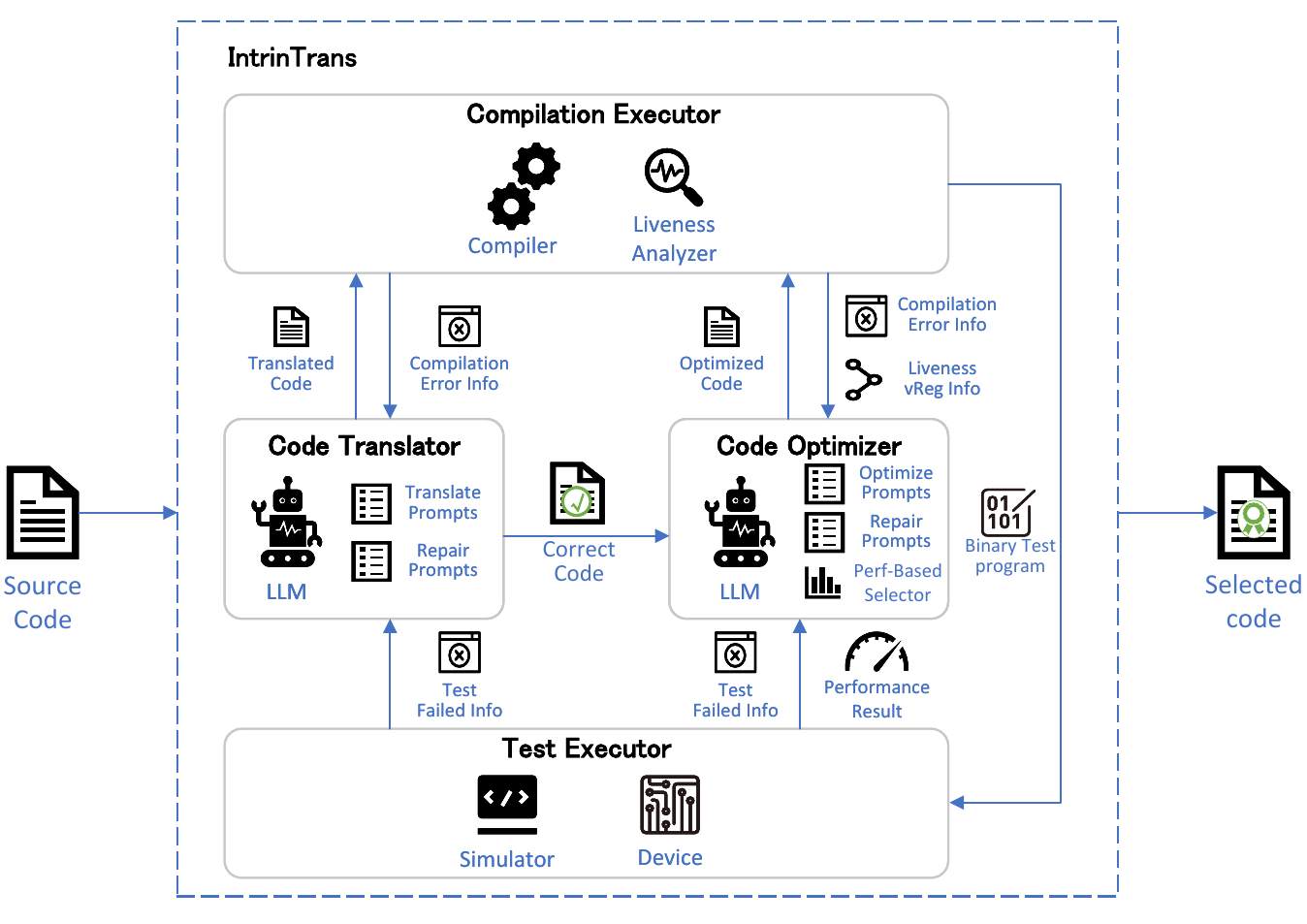}
  \caption{Design overview of IntrinTrans}
  \label{fig:IntrinTrans}
\end{figure}

% IntrinTrans 是一个 其他架构 Intrinsic code 到 RVV Intrinsic code 翻译工具，使用LLM-based多智能体翻译向量化代码到RISC-V架构上. 该工具接受任意Intrinsic 实现的向量化函数实现及其测试代码，接收到输入后，IntrinTrans使用多种智能体的能力生成RVV Intrinsic，执行编译-测试流程验证其正确性，并基于活跃变量分析判断寄存器使用情况，尝试生成能够充分利用硬件资源的RVV Intrinsic代码，并依据性能测试选取性能最佳的实现。
In this paper, we propose \texttt{IntrinTrans}, a novel translation framework designed to convert intrinsic-based vectorized code into the RVV intrinsics. As illustrated in Figure~\ref{fig:IntrinTrans}, our framework is architected around two core components: \textbf{LLM-based agents} for the primary code generation and an \textbf{automated feedback pipeline} for iterative validation and optimization. Given any vectorized function implementation written with architecture-specific intrinsics and its corresponding test code, \texttt{IntrinTrans} coordinates multiple specialized agents to generate RVV intrinsic implementations. It then performs a compile-and-test validation pipeline to ensure functional correctness. Furthermore, by conducting live variable analysis, the system evaluates register utilization and attempts to generate RVV intrinsic code that fully exploits available hardware resources. Finally, \texttt{IntrinTrans} conducts performance benchmarking and selects the implementation that achieves the best runtime efficiency.

\subsection{LLM-based agents}

% LLM agents 作为开发人员和大语言模型之间的intermediaries，抽象了与LLM API交互的复杂性。我们在IntrinTrans中引入了RVV-code Translator, Compilation Executor, Test Executor, Optimizer四个智能体，They maintain prompts and context windows, 并相互通信, 从而完成代码翻译任务。

% The RVV-code Translator

% The Compilation Executor

% The Test Executor
% Test Executor 可以将代码运行在真实的硬件环境中，从而从真实环境中获取相应的反馈。Test Executor 首先在本地对翻译完成的代码进行交叉编译（通过 VecIntrinBench 的构建系统），进行互动执行代码并获取相应的

% The Optimizer

LLM agents serve as intermediaries between developers and LLMs, abstracting the complexities of interacting with LLM APIs. We introduce four distinct agents in \texttt{IntrinTrans}: the RVV Code Translator, the Compilation Executor, the Test Executor, and the RVV Code Optimizer. These agents maintain their own dedicated prompts and context windows, communicating and collaborating with each other to accomplish the end-to-end code translation task.

\subsubsection{RVV Code Translator}
The RVV Code Translator is responsible for the initial translation of source intrinsic code into RVV-intrinsic representation. It interacts with an LLM using a set of carefully crafted translation and repair prompts. This agent operates in a closed loop, receiving feedback in the form of compilation errors and test failures from other agents, which enables it to iteratively refine and correct the generated code until a functionally correct version is produced.

\subsubsection{Compilation Executor}
The Compilation Executor takes the translated code and attempts to compile it. Beyond simple compilation, it performs static analysis, such as liveness analysis for virtual registers. It provides two critical feedback streams: it sends compilation error information back to the RVV Code Translator for error correction, and it forwards successfully compiled code along with analysis metadata to the RVV Code Optimizer to guide performance improvements.

\subsubsection{Test Executor}
The Test Executor provides dynamic feedback by running the compiled code in a realistic environment, such as on actual hardware or in a simulator. It first cross-compiles the translated code locally and then executes it interactively to capture runtime feedback. This process yields crucial information on execution failures or performance results, which is fed back to the RVV Code Translator and RVV Code Optimizer agents for further refinement.

\subsubsection{RVV Code Optimizer}
The Code Optimizer is tasked with enhancing the performance of the correctly translated intrinsic code. It synthesizes information from all other agents: the correct code from the Translator, static analysis data from the Compilation Executor, and runtime performance results from the Test Executor. Using this holistic view, it guides an LLM with specialized optimization prompts and employs a performance-based selector to explore and identify more efficient code variants, ultimately selecting the optimal implementation.

% \begin{table*}[!t]
% \caption{Evaluated LLMs in Our Experiments}
% \label{LLMs}
% \centering
% \setlength{\tabcolsep}{4pt}
% \begin{tabular}{@{}lll lll lll@{}}
%     \toprule
%     \textbf{Model} & \textbf{Tag} & \textbf{Org.} &
%     \textbf{Model} & \textbf{Tag} & \textbf{Org.} &
%     \textbf{Model} & \textbf{Tag} & \textbf{Org.} \\
%     \midrule
%     Claude-sonnet-4.5 & 20250929 & Anthropic & DeepSeek-V3 & 0324 & DeepSeek & Qwen3 Max & {\color[HTML]{09090B} qwen3-max} & Alibaba \\
%     Claude-opus-4 & 20250514 & Anthropic & DeepSeek-R1 & 0528 & DeepSeek & Gemini 2.5 Flash & - & Google \\
%     Claude-sonnet-4 & 20250514 & Anthropic & Grok 4 Fast & {\color[HTML]{09090B} non-reasoning} & xAI & Gemini 2.5 Pro & - & Google \\
%     Claude-3-5-Haiku & 20241022 & Anthropic & Grok 4 & 0709 & xAI & GPT-4o & 2024-11-20 & OpenAI \\
%     Claude-3-5-Sonnet & 20241022 & Anthropic & Grok 3 & - & xAI & GPT-5 Codex & - & OpenAI \\
%     DeepSeek-V3.2 & Exp & DeepSeek & Qwen3 Coder & {\color[HTML]{09090B} 480B-A35B-Instruct} & Alibaba & GPT-5 & {\color[HTML]{09090B} 2025-08-07} & OpenAI \\
%     DeepSeek-V3.1 & {\color[HTML]{09090B} Terminus} & DeepSeek & Qwen3 Coder Plus & {\color[HTML]{09090B} qwen3-coder-plus} & Alibaba & GPT-o3 & 2025-04-16 & OpenAI \\
%     \bottomrule
% \end{tabular}
% \end{table*}

\subsection{Automated Feedback Pipeline}

% 为了生成正确且高效的RVV intrinsic code, 我们通过a finite state machine指定多智能体之间的转换和通信，构成了保证正确性的Register-Usage-and-Speedup feedback loop和优化性能的Register-Usage-and-Speedup feedback loop。

% 在 Compile-and-Test-Error feedback loop 中， the compilation executor对 the RVV-code Translator的翻译产物(RVV Intrinisic code)进行编译，判断代码是否符合RVV Intrinsic的声明和调用方法，如果出现编译错误，如\texttt{use of undeclared identifier} 或 \texttt{no matching functions}等，则将编译报错信息反馈给the RVV-code Translator，要求其根据错误信息修复代码，如此循环，直到代码被成功编译，则会进一步与功能测试进行链接，生成用于评估正确性的可执行程序，交由The Test Executor。The Test Executor会在RISC-V不同\texttt{VLEN}的设备上执行测试，全面的评估代码正确性，如果任何设备上的任何一个测试用例失败，则将测试用例和测试失败情况反馈给the RVV-code Translator，要求其修复代码，如此循环直至通过测试，或达到给定的尝试次数阈值。

% 在 Register-Usage-and-Speedup feedback loop 中，the compilation executor 会在执行编译检查的基础上额外进行RISC-V Vector变量的活跃性分析，并据此计算寄存器使用量。

% 我们在 LLVM 编译器的基础上设计了针对RISC-V Vector变量的针对性分析工具。其本质是分析所有basic block中活跃的vector value，再根据其寄存器分组(LMUL)配置计算出占用的寄存器个数，据此分析得出寄存器使用的峰值，称之为寄存器压力（\text{vReg_Pressure}）, 它反映了向量代码对于硬件寄存器资源的需求程度。寄存器压力应当接近但不超过实际硬件的寄存器数量，否则会浪费硬件资源或造成register spilling，进而影响性能。

% 分析的过程主要是分析活跃的向量变量，最终选出所有位置的最大使用数即为需要的向量寄存器资源数。
% 活跃变量分析的首先要为每个vector expression定义 USE，DEF，IN，OUT 集合
% 其中
% - USE 表示当前 expression 使用的向量 value集合
% - DEF 表示当前 expression 定义的向量 value集合
% - IN 当前 expression 入口活跃的向量 value集合，包含在执行该exp之前已经处于活跃状态的向量value集合
% - OUT 表示当前 expression 出口活跃的向量 value集合，包含在执行该exp之后仍然处于活跃状态的向量value集合

% 之后，自然可以得到每个expression的 USE 集合和 DEF集合，再根据活跃变量分析的数据流方程，采用后向迭代的方式求解每个expression的IN和OUT集合，重复此过程直到所有集合不再变化为止。方程中Successor(i)表示expression i的后继表达式。

To generate correct and efficient RVV intrinsic code, we orchestrate the transitions and communications between the agent system using a finite state machine. This framework is architected around two primary feedback loops: a Compile-and-Test-Error feedback pipeline to ensure functional correctness, and a Register-Usage-and-Speedup feedback loop to optimize performance.

\subsubsection{Compile-and-Test-Error Feedback Pipeline for Correctness}
In this iterative pipeline, the Compilation Executor first receives the translated intrinsic code from the RVV Code Translator. It then attempts to start the compilation process, which statically verifies whether the code conforms to the declaration and invocation conventions of RVV intrinsics. If compilation errors are detected, such as the use of an undeclared identifier or no matching functions, the specific error messages are relayed back to the RVV Code Translator. This feedback prompts the agent to repair the code based on the provided error context. This iterative cycle of translation and compilation continues until the code is successfully compiled.

Upon successful compilation, the artifact is linked with a functional test suite to generate an executable program, which is then dispatched to the Test Executor. The Test Executor comprehensively evaluates the program's correctness by running it on real-world RISC-V devices with different \texttt{VLEN} values. Should any test case fail on any device, a detailed report including the specific test case and failure conditions is fed back to the RVV Code Translator to initiate another repair cycle. This cycle persists until the code passes all tests or a predefined attempt threshold is exceeded.

\subsubsection{Register-Usage-and-Speedup Feedback Pipeline for Optimization}
In this iterative pipeline, the Compilation Executor augments its standard compilation check with a specialized liveness analysis for RISC-V vector variables to calculate register usage.

For this purpose, we have designed a targeted analysis tool built upon the LLVM compiler infrastructure~\cite{llvm}. Its fundamental task is to analyze the live vector values across all basic blocks. Based on the live values and their corresponding register grouping configuration (\texttt{LMUL}), the tool calculates the number of physical registers occupied. This analysis yields the peak register usage, which we term \textit{vector register pressure}. This metric reflects the vector code's demand on hardware register resources. Optimal performance is often achieved when the register pressure is high but does not exceed the number of available hardware registers, as this avoids both resource underutilization and performance-degrading register spilling.

The analysis process identifies the maximum number of concurrently live vector variables at any program point, which corresponds to the required vector register resources. To formalize this, we define four sets for each vector expression $i$: 1) $USE(i)$ is the set of vector values used in expression $i$; 2) $DEF(i)$ denotes the set of vector values defined (written to) by expression $i$; 3) $IN(i)$ represents the set of vector values that are live at the entry point of expression $i$; 4) $OUT(i)$ is defined as the set of vector values that are live at the exit point of expression $i$.

The $USE$ and $DEF$ sets for each expression are determined by direct inspection. Subsequently, the $IN$ and $OUT$ sets are computed by solving the following data-flow equations via a backward iterative analysis, where $Successor(i)$ denotes the set of succeeding expressions for expression $i$:

\begin{equation}
IN(i) = (OUT(i) - DEF(i)) \cup USE(i)
\label{eq-in}
\end{equation}

\begin{equation}
OUT(i) = \bigcup_{j \in Successor(i)} IN(j)
\label{eq-out}
\end{equation}

% 数据流分析得到的IN 和 OUT 集合的并集表示了当前 expression 位置活跃的向量变量，再根据每个变量的LMUL配置就可以通过公式X求出基本块内的向量寄存器使用数量的最大值，如方程TODO标号所示。
The union of the $IN$ and $OUT$ sets for a given expression represents the complete set of live vector variables at that program point. The number of physical registers required at this point can be calculated by summing the \texttt{LMUL} configuration of each live variable. The overall vector register pressure for the code block is then determined by the maximum value of this sum across all expressions, as quantified by Eq.~\ref{eq-max}.

\begin{equation}
\text{vReg}_{\scriptscriptstyle \text{Pressure}} = \max_{i} \left(\sum_{v \in IN[i] \cup OUT[i]}  LMUL(v) \right)
\label{eq-max}
\end{equation}

% the compilation executor 会将寄存器压力反馈给 the Optimizer, 后者会根据尝试采用不同的LMUL策略或使用循坏展开寄存更合理的使用RISC-V Vector的32个Vector Register，进一步优化The RVV-code Translator的翻译产物。与Compile-and-Test-Error feedback loop类似，Register-Usage-and-Speedup feedback loop 中的 Optimizer 也会和the Test Executor 交互，但此时the Test Executor在验证代码正确性的基础上还会反馈性能数据，此性能数据是优化后相对于优化前(即翻译产物)的加速比。the Optimizer会记录每一个正确编译且通过了测试的代码及其性能数据，最终选取性能最佳的实现作为交付给用户的输出结果。
The Compilation Executor reports this calculated register pressure to the RVV Code Optimizer. The Optimizer then uses this feedback to further refine the output from the RVV Code Translator. It may employ various strategies, such as experimenting with different \texttt{LMUL} policies or applying loop unrolling, to make more judicious use of the 32 physical vector registers available in the RISC-V architecture.

Similar to the Compile-and-Test-Error pipeline, the Code Optimizer in the Register-Usage-and-Speedup feedback pipeline also interacts with the Test Executor. However, in this phase, the Test Executor's role is twofold: in addition to verifying the functional correctness of the optimized code, it also provides crucial performance data. This data is typically presented as a speedup ratio, comparing the performance of the optimized version against the initial, unoptimized (but correct) translated artifact. The RVV Code Optimizer systematically records every code variant that compiles correctly and passes all functional tests, along with its corresponding performance data. Finally, it selects the implementation exhibiting the best performance as the final output to be delivered to the user.

\begin{table}
\centering
\caption{Evaluated LLMs in Our Experiments.}
\label{LLMs}
\resizebox{\columnwidth}{!}{%
\begin{tabular}{llll}
\hline
\textbf{Model}    & \textbf{Tag} & \textbf{Org.} & \textbf{Reasoning} \\ \hline
Claude-sonnet-4.5 & 20250929     & Anthropic     & Enable/Disable     \\
Grok 4            & 0709         & xAI           & Enable             \\
Gemini 2.5 Pro    & -            & Google        & Enable             \\
GPT-5 Codex       & -            & OpenAI        & Enable             \\
DeepSeek-V3.2     & Exp          & DeepSeek      & Enable/Disable     \\
DeepSeek-R1       & 0528          & DeepSeek     & Enable     \\
Qwen3 Coder Plus  & -            & Alibaba       & Disable            \\ \hline
\end{tabular}%
}
\end{table}

\section{Experiment}
\label{sec:exp}
% 我们设计实验来探索以下研究问题：
We designed experiments to explore the following research questions:

\begin{itemize}
\item RQ1: How correct and efficient is \texttt{IntrinTrans} in translating code across architectures to RISC-V Vector Intrinsic?
% IntrinTrans 在将代码跨架构翻译到RISC-V Vector Intrinsic任务上的正确性和效率如何？
\item RQ2: What is the performance of RISC-V Vector Intrinsic code generated by \texttt{IntrinTrans}?
% IntrinTrans 生成的有效RISC-V Vector Intrinsic代码的性能如何？
\end{itemize}

\subsection{Benchmark and Evaluation Metrics}

% 我们使用我们提出的VecIntrinBench作为Benchmark，并沿用其中的pass rate和speedup作为evaluation metrics。 其中，The \textbf{pass rate} indicates the effectiveness of the translation method, 而 speedup is a commonly used indicator to evaluate the optimization effect of an algorithm, and is used to measure the optimization effect of each valid translation product. 额外的，为了更好地评估基于反馈迭代的智能体翻译效果，我们在此引入\textbf{efficiency score}作为新的evaluation metrics.

We employ our proposed VecIntrinBench as the benchmark and adopt its pass rate and speedup as evaluation metrics. The pass rate quantifies the effectiveness of the translation method. In contrast, speedup is a widely used metric for assessing algorithmic optimization and is used here to measure the optimization achieved by each valid translation output. In addition, to better evaluate the translation performance of the feedback-iterative agent, we introduce an efficiency score as a new evaluation metric.

% 如公式\ref{eq-efficiency}所示,Efficiency score 量化了翻译方法通过反复迭代最终完成代码翻译任务的效率。对所有的case的\texttt{Efficiency score}_i求和就得到了翻译方法的Efficiency score，该值越高则表示方法需要的迭代尝试次数越少，通常意味着其效率更高。
As shown in the formula \ref{eq-efficiency}, the \textbf{efficiency score} quantifies the efficiency of the translation method in completing the code translation task through repeated iterations. A higher value indicates that the method requires fewer iterations, which generally indicates higher efficiency.

\begin{equation}
\text{efficiency score} = \sum_{i\in Cases} \left( \frac{1+\text{upLimit} - attempts}{\text{upLimit}}  \right)
\label{eq-efficiency}
\end{equation}

\subsection{Experiment Setup}

We evaluated 9 LLMs from six leading organizations on \texttt{IntrinTrans}, encompassing both state-of-the-art LLMs (as of the time our experiments were conducted) and classic LLMs. Table \ref{LLMs} lists the model names, labels, and their organizations. All evaluated LLMs were accessed via an aggregation platform that invoked the corresponding official APIs. We set the sampling temperature to 0.2, appropriate for code generation tasks, allowed \texttt{IntrinTrans} to perform up to 10 translation iterations and an optimization iteration, then computed the pass rate, efficiency score, and speedup. We used the GCC 14.2 compiler to build both the native code and valid translated code with the same options (\texttt{-march=rv64gcv -O3}) for the RVV architecture, evaluated correctness on QEMU at \texttt{VLEN=128-bit} and \texttt{VLEN=256-bit}, and measured performance on a SpacemiT Muse-Pi, which is a RISC‑V board based on SpacemiT M1 CPU and a VLEN of 256 bits.

% 我们在IntrinTrans上分别评估了来自6家领先组织的9个大语言模型，这些模型既包括最先进的LLMs(直到我们开展使用),也包括经典的LLMs, 表\ref{LLMs}列出了模型名称，标签及其组织，其中，位于表中最后一列的GPT‑5-Codex, GPT-5, GPT-o3，Gemini-2.5-Pro 和 DeepSeek-R1 具备并启用了reasoning能力。所有评估的LLM使用聚合平台调用相应的官方API访问。我们将采样温度设置为适合代码生成任务的0.2，允许IntrinTrans 进行至多10次翻译迭代和10次优化迭代，并计算pass rate，efficiency score和speedup。我们使用GCC 14.2编译器将native code和有效的translated code以同样的编译选项(-march=rv64gcv -O3)编译到RISC-V Vector架构上，在QEMU上评估分别其在VLEN=128-bit和VLEN=256-bit时的正确性，并在运行 Linux 6.6.36 的 SpacemiT MUSE Card 开发板上评估性能，这是一款采用SpacemiT M1 CPU的RISC-V开发板，它的VLEN是256-bit。

\begin{table}[]
\caption{Pass Rate, Average Iterations, and Efficiency Score of Different Methods on VecIntrinBench}
\label{pass}
\resizebox{\columnwidth}{!}{%
\begin{tabular}{lccc}
\hline
\textbf{LLM/Rule-based Methods} & \multicolumn{1}{l}{\textbf{Pass Rate}} & \multicolumn{1}{l}{\textbf{Average Iterations}} & \multicolumn{1}{l}{\textbf{Efficiency Score}} \\ \hline
\textbf{GPT-5-Codex} & 100.0\% & 1.8 & 31.7 \\
\textbf{Claude-Sonnet-4.5} & 100.0\% & 1.8 & 30.9 \\
\textbf{Gemini-2.5-pro} & 97.1\% & 2.6 & 28.5 \\
\textbf{Grok-4} & 94.1\% & 1.8 & 31.5 \\
\textbf{Claude-Sonnet-4.5(non-reasoning)} & 88.2\% & 2.0 & 27.7 \\
\textbf{DeepSeek-R1} & 70.6\% & 2.0 & 21.9 \\
\textbf{Qwen3-Coder-Plus} & 64.7\% & 2.6 & 21.1 \\
\textbf{DeepSeek-v3.2} & 61.8\% & 3.0 & 17.9 \\
\textbf{Neon2RVV} & 52\% & - & - \\
\textbf{DeepSeek-v3.2 (non-reasoning)} & 47.1\% & 2.4 & 16.1 \\ \hline
\end{tabular}%
}
\end{table}

\subsection{RQ1: Correctness Evaluation}

% 我们将34个Arm Neon Inrtinsic向量化的函数交由 IntrinTrans 执行面向 RISC-V Vector Intrinsic的代码翻译任务，The Compile-and-Test-Error feedback loop的迭代上限被设置为10次，在有限次迭代中产生能够编译出二进制程序并通过全部测试的代码视为完成任务。我们使用 test pass rate 作为评价正确性的指标并使用平均迭代次数和efficiency score综合评价效率。

% 如图\ref{fig:RQ1}所示的实验结果表明，基于不同的LLMs，IntrinTrans翻译产物的测试通过率在24%到100%之间。LLMs对代码翻译的正确性具有显著影响，先进的LLM通常具有更好地翻译正确性，且reasoning model在代码翻译任务中具有明显的优势，可能的原因在于其能够更好地理解和利用IntrinTrans给出的错误反馈信息。

We assigned 50 vectorized functions written in Arm Neon Intrinsic to \texttt{IntrinTrans} for code translation to RISC-V Vector Intrinsics. We used the test pass rate as a metric for evaluation. Table~\ref{pass} showed that the test pass rates of \texttt{IntrinTrans}'s translations ranged from 47\% to 100\% based on different LLMs, while the pass rate of rule-based \texttt{Neon2RVV} is 52\%. Advanced LLMs significantly impacted the accuracy of code translation, and the reasoning model also demonstrated a clear advantage in code translation, likely due to its ability to better understand and utilize the error feedback provided by \texttt{IntrinTrans}.

% 为了探究IntrinTrans的feedback loop在何种程度上影响LLMs在代码翻译场景下的效果，我们记录了所有任务的反馈循环执行次数，如表所示，% 我们从9个LLMs的平均迭代次数为1.9~5.4次。在部分短小的任务中，迭代次数仅为1，这说明在简单算法的翻译任务上，LLMs可以在提示词的帮助下直接生成正确的RISC-V Vector Intrinsic，无需反馈循环。但对于更复杂的任务或non-SOTA的LLMs而言，IntrinTrans的错误反馈机制对于面向RISC-V架构的代码翻译任务而言仍是必要的。
To explore the extent to which \texttt{IntrinTrans}'s feedback pipeline affects the effectiveness of LLMs in the code translation scenario, we recorded the number of feedback loop executions for all tasks, as shown in Table~\ref{pass}. The average number of iterations ranged from 1.8 to 3, and some short cases required only a single iteration, while up to nine feedback iterations were needed to translate the complicated code accurately. This demonstrates that for simple algorithmic translation tasks, LLMs can directly generate correct RISC-V vector intrinsics with the help of prompts, without the need for a feedback loop. However, for more complex tasks or non-SOTA LLMs, the error feedback mechanism of \texttt{IntrinTrans} remains essential for code translation tasks targeting the RISC-V architecture.

\subsection{RQ2: Performance Evaluation} 

\begin{figure*}[t]
  \centering
  \includegraphics[width=0.80\linewidth]{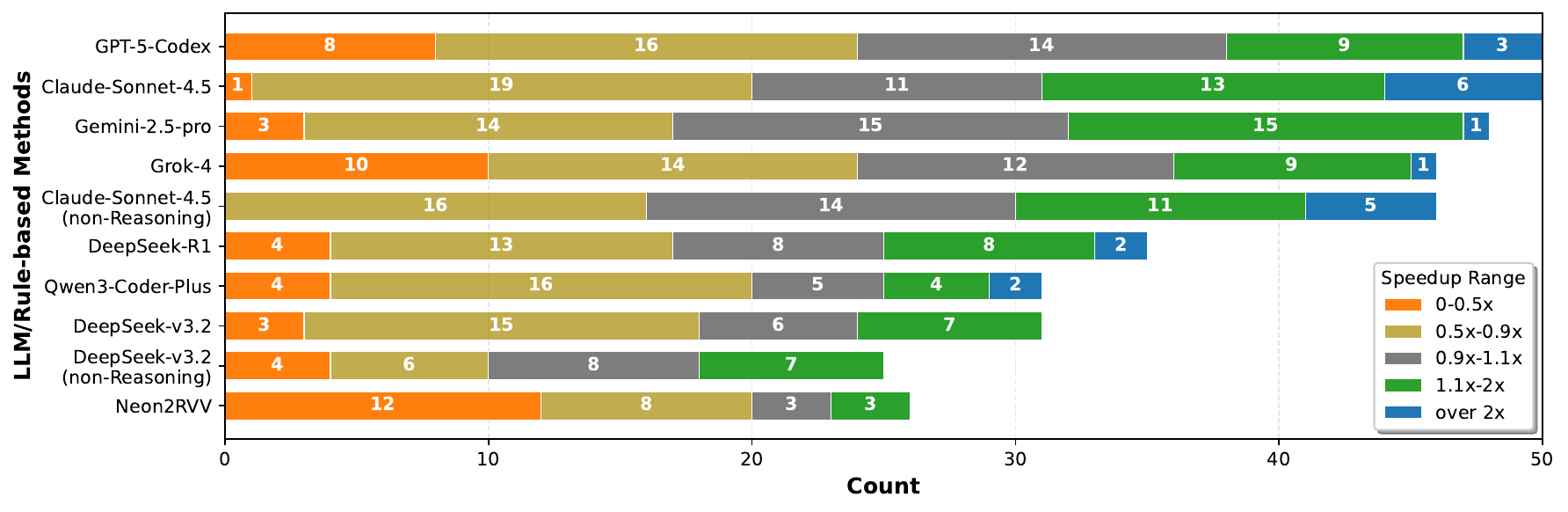}
  \caption{Speedup Distribution of Translated Code (Native Code as Baseline)}
  \label{RQ3}
\end{figure*}

% 我们在本节说明 IntrinTrans 的翻译产物和人类专家编写的代码(native code)性能差异，并探究基于活跃变量的Register-Usage-and-Speedup feedback的有效性。以 native code 作为性能基线，IntrinTrans在不同LLMs的翻译产物的性能测试执行时间与之做对比，可以通过得到各个测试用例的加速比。总的来说，在所有479个成功翻译的任务组合中，有132个任务的加速比大于1.1倍，占27.6%。加速比的分布区间如\ref{RQ3}所示。图中橙色表示翻译产物的性能与native code相差巨大，而黄色则说明加速比在0.5到0.9倍之间，说明翻译产物仍有优化空间，这可能是由于当前的优化反馈仅提供活跃变量和寄存器使用情况的信息，而没有考虑访存局部性等其他优化因素导致的。绿色表示翻译产物的性能优于native code，而蓝色表示性能是native code的一倍以上，最大的加速比是基于Gemini 2.5 Pro翻译\texttt{h2v1\_upsample}任务的5.93倍。这说明IntrinTrans在向量化领域具有一定的性能优势，也在一定程度上表明native code的质量不佳。进一步观察，翻译产物存在性能优势的case多来自于Libjpeg-turbo仓库，其实现欠佳的原因可能是在benchmark采集时，其RISC-V Vector的实现仍处于under review的pull request状态，尚未被合入主线仓库，质量未经全面评估。
In this section, we compare the performance of translated code with that of native code written by human experts and explore the effectiveness of the register-usage and speedup feedback pipeline based on active variables. Using native code as a performance baseline, we compare the performance test execution time of \texttt{IntrinTrans}'s translated code across different LLMs, yielding speedup ratios for each test case. Overall, the performance of the optimized translation code ($0.85\times$ to $1.28\times$) is close to that of the native intrinsic performance, exceeding the rule-based Neon2RVV($0.63\times$). As shown in table~\ref{ablation}, compared to before the introduction of register-usage-and-speedup feedback, the performance of these tasks improved by an average of 9.19\% to 34.87\%. The distribution of speedup ratios is shown in Figure~\ref{RQ3}. Orange indicates significant performance differences between the translated code and native code, while yellow indicates speedup ratios between $0.5\times$ and $0.9\times$, indicating potential for optimization. This may be due to the current optimization feedback only providing information on liveness variables and register usage, without considering other optimization factors such as cache and locality of reference. Green indicates that the translated code outperforms the native code, while blue indicates performance is more than double that of the native code. A higher speedup indicates that \texttt{IntrinTrans} has a performance advantage in vectorization, which usually implies poor native code quality.

\begin{table}[]
\caption{Ablation Study:  The Role of Optimization Feedback Based on Liveness Analysis}
\label{ablation}
\resizebox{\columnwidth}{!}{%
\begin{tabular}{lccc}
\hline
\textbf{LLM/Rule-based Methods} & \multicolumn{1}{l}{\textbf{Tranlated Speedup}} & \multicolumn{1}{l}{\textbf{Optimized Speedup}} & \multicolumn{1}{l}{\textbf{Improvement}} \\ \hline
\textbf{GPT-5-Codex} & 0.84 & 0.98 & 17.09\% \\
\textbf{Claude-Sonnet-4.5} & 1.01 & 1.28 & 25.99\% \\
\textbf{Gemini-2.5-pro} & 0.97 & 1.10 & 13.78\% \\
\textbf{Grok-4} & 0.68 & 0.91 & 34.87\% \\
\textbf{Claude-Sonnet-4.5 (non-reasoning)} & 0.97 & 1.20 & 24.00\% \\
\textbf{DeepSeek-R1} & 0.89 & 0.97 & 9.50\% \\
\textbf{Qwen3-Coder-Plus} & 0.78 & 0.88 & 13.22\% \\
\textbf{DeepSeek-v3.2} & 0.76 & 0.85 & 11.84\% \\
\textbf{DeepSeek-v3.2 (non-reasoning)} & 0.81 & 0.89 & 9.19\% \\
\textbf{Neon2RVV} & 0.63 & - & - \\ \hline
\end{tabular}%
}
\end{table}

\section{Conclusion}

This work introduced \texttt{IntrinTrans}, a LLM agent framework designed to address the longstanding challenges of cross-architecture intrinsic translation, with a particular focus on the RVV extension. Distinct from conventional rule-based approaches that rely on static instruction mappings, \texttt{IntrinTrans} leverages the semantic reasoning and adaptive generalization capabilities of LLMs within a structured feedback loop, including translation, evaluation, and optimization. This design enables the system to reconcile the trade-off between semantic correctness and hardware-aware performance when translating intrinsic-level code.

We systematically evaluated \texttt{IntrinTrans} using previously open-sourced intrinsic code benchmarks. A comprehensive study of nine major language models demonstrates that \texttt{IntrinTrans} can generate functionally correct and high-performance RVV code from Arm Neon intrinsic functions, thus validating the feasibility of using LLMs for automated cross-ISA code translation.

The findings of this study point toward a new paradigm in compiler-assisted code generation—one that integrates LLM-based semantic reasoning with architecture-specific optimization. Future research will focus on reinforcing optimization-aware reasoning within the agent workflow, extending benchmark diversity to broader RISC-V extensions, and exploring hybrid fine-tuning strategies for domain-adaptive intrinsic translation. \texttt{IntrinTrans} will be open-sourced once this paper is accepted, thereby helping to develop the RISC-V software ecosystem more effectively.

%%%%%%%%%%%%%%%%%%%%%%%%%%%%%%%%%%%%%%%%%%%%%%%%%%%%%%%%%%%%%%%%%%%%%%%%

%%% The next two lines define, first, the bibliography style to be 
%%% applied, and, second, the bibliography file to be used.

\bibliographystyle{IEEEtran}
\bibliography{sample}

%%%%%%%%%%%%%%%%%%%%%%%%%%%%%%%%%%%%%%%%%%%%%%%%%%%%%%%%%%%%%%%%%%%%%%%%

\end{document}